\documentclass[preprint,a4paper,showpacs,superscriptaddress]{revtex4}
\usepackage{graphicx}
\usepackage{bm}
\begin{document}
\title{Noise of Kondo dot with ac gate: Floquet-Green's function and Noncrossing Approximation Approach}
\author{B. H. Wu}
\email{bhwu@mail.sim.ac.cn} \affiliation{Max Planck Institute for
the Physics of Complex Systems, 38 Noethnizer Str. D-01087 Dresden,
Germany} \affiliation{State Key Laboratory of Functional Materials
for Informatics, Shanghai Institute of Microsystem and Information
Technology, Chinese Academy of Sciences, 865 Changning Road,
Shanghai 200050, People's Republic of China }
\author{J. C. Cao}
\affiliation{State Key Laboratory of Functional Materials for
Informatics, Shanghai Institute of Microsystem and Information
Technology, Chinese Academy of Sciences, 865 Changning Road,
Shanghai 200050, People's Republic of China }
\begin{abstract}
The transport properties of an ac-driving quantum dot in the Kondo
regime are studied by the  Floquet-Green's function method with
slave-boson infinite-$U$ noncrossing approximation. Our results show
that the Kondo peak of the local density of states is robust against
weak ac gate modulation. Significant suppression of the Kondo peak
can be observed when the ac gate field becomes strong. The
photon-assisted noise of Kondo resonance as a function of dc voltage
does not show singularities which are expected for noninteracting
resonant quantum dot. These findings suggest that one may make use
of the photon-assisted noise measurement to tell apart whether the
resonant transport is via noninteracting resonance or
strongly-correlated Kondo resonance.
\end{abstract}
\maketitle

\section{introduction}
The Kondo effect is a paradigm of strong correlation in condensed
matter physics.\cite{Hewson} This effect was first discovered in
metals with magnetic impurity where the resistivity enhances with
decreasing temperature below a characteristic temperature known as
Kondo temperature $T_K$. Such anomaly behavior was theoretically
explained by J. Kondo, hence the name ``Kondo effect", as the
exchange interaction of the itinerant electrons with a localized
spin state.\cite{Hewson} Soon, it was realized that the same
correlation  can dominate the low temperature transport properties
of quantum dots with strong Coulomb
interaction.\cite{PRL611768,JETP47452} Contrary to metals, the Kondo
effect leads to the enhancement of conductivity of the quantum dots.
The advance of nanotechnology has experimentally demonstrated the
Kondo effect in artificial impurity systems (semiconductor quantum
dots\cite{Nature3911569,Science281540}, carbon
nanotube\cite{Nature408342} or molecular
conductors\cite{Nature417722,Nature417725}). The observation of
Kondo effect in these artificial quantum impurities has provided us
the opportunity to study correlation phenomena by tuning the
relevant parameters or driving the system out of equilibrium by
either static or alternate magnetic or electric field. However,
comparing with the great achievement of experiments, the theoretical
understanding of Kondo effect, especially at nonequilibrium, is far
from adequate. This difficulty is due to the interplay of strong
Coulomb interaction and nonequilibrium which makes the exact
solution of the Kondo problem in nonequilibrium impossible. For the
theoretical studies of Kondo effect, several techniques, such as
equation of motion of Green's function\cite{PRB73125338},
bosonization technique\cite{PRB362036,RMP59845,PRB4911040}, and
numerical renormalization group method\cite{RMP80395}, with their
own advantages and limitations are developed. Up to now, a large
literature has been devoted to study the transport properties of
Kondo systems. Most of these studies focused on the spectral
density, linear or differential conductance and the current.

Nowadays, there is increasing interest in the study of the
fluctuation of current, i.e. the current noise\cite{PR3361}, in the
Kondo regime. The reasons for these studies are roughly two folds.
On one hand, current noise measurement is expected to expose
information about the quantum nature of electron transport in Kondo
regime.  For Kondo dot, the Kondo correlation is quite sensitive to
the external electric or magnetic field. It is thus difficult to get
some key information such as the Kondo temperature or the spectral
density by direct experimental measurement. New setups or novel
characteristic tools are needed to get the desired information. The
current noise of Kondo dot has shown its potential to provide
information which is beyond the reach of traditional transport
measurement. For instance, Meir and Golub studied the shot noise of
Kondo dot by several complementary approaches\cite{PRL88116802}.
They proposed to estimate  $T_K$ by measuring the noise behavior of
Kondo dot. In addition, Sindel \emph{et al.} pointed out that the
current noise can give an estimate of the Kondo resonance in the
equilibrium spectral density.\cite{PRL94196602} On the other hand,
as a signature of many-particle correlation, the current noise in
Kondo regime behaves quite differently from that of systems in
single particle picture. The recent study of fractional charge is
such an example. It is suggested that the noise measurement in Kondo
regime can lead to non-integer effective
charge\cite{PRL97086601,PRB75155313,PRB77R241303}. Moreover, the
recent experimental measurement reveals surprisingly high current
noise of carbon nanotube in the Kondo regime where the conductance
is close to the Landauer conductance quantum.\cite{NP5208} This is
in great contrast to the predictions based on the noninteracting
picture where the noise is expected to be negligible.

In this study, we show another example where the noise behavior of
ac-driving quantum dots in Kondo regime differs drastically from
that of noninteracting dots. For non-interacting quantum conductors,
Lesovik and Levitov (LL) have predicted that the noise should show
singularities at integers of $eV/\hbar\Omega$, where $V$ is the dc
bias, $\Omega$ is the ac frequency.\cite{PRL72538} Soon, this
prediction was experimentally verified by measuring the noise power
spectrum of a tunnel junction driven by microwave field.
\cite{PRL802437} Further theoretical studies, which in principle are
still in single-particle pictures, showed that the singularity is
robust against disorder effect and remains distinct at weak ac
field\cite{PRL91036804} (actually, the singularity also depends on
the ratio of $eV_{ac}/\hbar\Omega$ where $V_{ac}$ is the ac
strength). None of these studies included correlation effect due to
the Coulomb interaction in the conductor. When there is finite
Coulomb interaction in the conductor, the noise behavior is expected
to show different behavior since the electrons are correlated and
the single-particle picture is no-longer valid. Guigou \emph{et al.}
studied the noise in an ac-driving Luttinger
liquid.\cite{PRB76045104} Their results showed that the sharp
features in $dS/dV$ is smoothed by the presence of Coulomb
interaction. Since the electron transport in Kondo regime is
dominated by the strongly correlated Kondo resonance, an immediate
question one might ask is what happens to the photon-assisted noise
singularities in the Kondo regime. There exists some literatures
which studied the transport properties of the ac Kondo
regime.\cite{PRL744907,PRL76487,PRB56R15521,PRL83808,PRB612146,PRL814688,PRL815394}
Most of them focus on the conductance or density of states of an
ac-driving Kondo dot . There have hitherto been few attempts to
study the noise properties in ac-driving Kondo
dot\cite{PRB56R15521,PRB77233307,EPL8268004}. Recently, the noise of
Kondo dot with ac gate modulation was studied\cite{PRB77233307} by
present authors by combining the Floquet theorem and the slave boson
mean field approximation. This mean-field approximation is valid
when the Kondo temperature defines largest energy scale. The
numerical results showed that the photon-assisted noise of Kondo dot
shows no singularity at zero temperature and very low bias voltage.
This smooth noise behavior was in great contrast to the predictions
in single particle picture. Limited by the validity of slave-boson
mean field approximation, it remains unclear whether such
singularity can appear in the Kondo regime when the Kondo
temperature does not define the largest energy scale.

It is the purpose of the present paper to investigate the
photon-assisted noise of Kondo dot beyond the slave boson mean field
approximation.  Our approach is based on the noncrossing
approximation
(NCA)\cite{PRB362036,PRB4913929,PRB4911040,PRB67245307}.  For
$U\rightarrow\infty$, where $U$ is the Coulomb interaction strength,
the NCA has been shown to give reliable results both in the
temperature regimes above and below $T_K$. The splitting of the
Kondo resonance by external dc bias can also be captured. However,
it fails to reproduce the Fermi liquid behavior at very low
temperature $T\ll T_K$ and gives non-physical results in the
mixed-valence region.\cite{PRB362036,JPSJ7416} A suitable remedy of
this shortcoming is possible as shown by Kroha \emph{et
al.}\cite{PRL79261,JPSJ7416}. The NCA has also been applied to the
time-dependent nonequilibrium transport through a Kondo
dot.\cite{PRL83808,PRB612146} Solution of the the time-dependent NCA
in the time domain has been proposed and
implemented.\cite{PRB4913929} In the present study,  we avoid to go
to the pathological regime of NCA\cite{PRB362036} and make use of
the periodicity of the ac field. We combine the Floquet
theorem\cite{PR406379,PR3951} and the infinite-$U$ NCA to
investigate the transport properties, in particular the current
noise properties of quantum dot with ac gate modulation. A suitable
Floquet-Green's function formalism is developed to describe the
dynamics of the system. The Floquet theorem can capture the
multi-photon process in a coherent non-perturbative way.  Our
results show that for low ac frequency and low ac strength, the
local density of states (LDOS) of the Kondo dot is almost
insensitive to the ac gate field. For high ac frequency and
strength, the Kondo peak can be significantly suppressed with
sidebands appearing in the density of states with distances to the
Kondo peak of multi-photon energy. Such suppression is attributed to
the decoherence induced by the photon-assisted
processes\cite{PRL83384}. Interestingly, the noise of quantum dots
in ac-Kondo regime shows dramatically different behavior from the ac
driving noninteracting dots. The noise as a function of dc bias
voltage does not show singularities in the Kondo regime as we change
the ac parameters from weak to strong values. One may thus utilize
the photon-assisted noise behavior to tell apart the single-particle
noninteracting resonant transport from the strongly correlated Kondo
resonance.

The present paper is organized as following. In Sec. II, we give the
details of the model Hamiltonian and the Floquet-Green's function
formalism with NCA. In Sec. III. numerical results of the LDOS and
noise properties of ac-driving Kondo dot are presented. For the sake
of comparison, results for non-interacting dot are given to show the
drastic different noise behaviors with or without many-body
correlations. In Sec. IV a conclusion is presented.

\section{Theoretical Formalism}

\subsection{Model Hamiltonian}
The quantum transport through quantum dot is widely described by the
single impurity Anderson model coupled to two ideal leads. The
system can be driven out of equilibrium by both dc bias and ac gate
field. Here, the ac field can be realized via a nearby ac gate which
modulates the dot energy level. To investigate the Coulomb effect on
the noise behavior, we are interested in two situations, i.e.
noninteracting dot without Coulomb interaction and the Kondo dot
with strong on-site Coulomb interaction.

For noninteracting dot, the system can be modeled by the single
impurity Anderson model with zero Coulomb strength ($U=0$) coupled
to two ideal leads. The Hamiltonian can be written as
\begin{eqnarray}
  H=\sum_{\alpha,k,\sigma}\epsilon_{k,\alpha,\sigma}
  c^\dag_{k,\alpha,\sigma}c_{k,\alpha,\sigma}
  +\sum_{k,\alpha,\sigma}(V_{k\alpha\sigma}c^\dag_{k,\alpha,\sigma}d_\sigma+h.c.)
  +\sum_\sigma \epsilon_d(t)d_\epsilon^\dag
  d_\sigma,
\end{eqnarray}
where $c^\dag_{k,\alpha,\sigma}$ is the creation operator of
$k$-electron with spin $\sigma$ in the $\alpha$th lead
($\alpha=L,R$), $\epsilon_{k,\alpha,\sigma}$ is the lead electron
energy. We assume a symmetric voltage drop across the dot.
$V_{k,\alpha,\sigma}$ denotes the coupling matrix element of the
lead electron and that in the dot. $d^\dag_\alpha$ is the creation
operator of electron in the dot with spin $\sigma$. Its energy level
can be modulated by an external gate voltage as
$\epsilon_d(t)=\epsilon_d^0+eV_g\cos(\Omega t)$. Here
spin-degeneracy is implied.

Usually, the simple assumption of zero Coulomb interaction can not
correctly capture the electron dynamics in nanostructures. For
quantum dot with small-size, the charging energy can be much larger
than the other energy scales. In these situation, we can
approximately take the limit $U\rightarrow\infty$. For infinite $U$,
the dot state can contain either zero or one electron. Double
occupancy may be neglected. A reliable tool to give a description of
quantum dots with level deep below the Fermi energy is the slave
boson NCA\cite{RMP59845,PRB4911040,PRB67245307}. The slave boson
technique introduces some auxiliary particles to eliminate the
non-quadratic term in the Hamiltonian by replacing the electron
operators by $ d^\dag_\sigma\rightarrow b^\dag f_\sigma$ and
$d_\sigma\rightarrow f^\dag_\sigma b_\sigma$, where $b$ annihilates
the empty state and $f_\sigma$ destroys a spin $\sigma$ single
occupation state. The infinite $U$ Hamiltonian in slave boson
language reads
\begin{eqnarray}
H=\sum_{\alpha k \sigma}\epsilon_{k\alpha
\sigma}(t)c^\dag_{k\alpha\sigma}c_{k\alpha\sigma}
+\sum_{k\alpha\sigma}(V_{k\alpha\sigma}c^\dag_{k\alpha\sigma}b^\dag
f_\sigma+h.c.) +\sum_\sigma \epsilon_{d}(t)f^\dag_\sigma f_\sigma.
\end{eqnarray}

Since the double occupancy is forbidden for $U\rightarrow \infty$,
these slave particle operators must satisfy the constraint condition
\begin{eqnarray}\label{Cost}
\sum_\sigma f^\dag_\sigma f_\sigma + b^\dag b =1.
\end{eqnarray}

\subsection{Floquet-Green's function formalism}

To study the transport properties out of equilibrium, we make use of
the nonequilibrium Green's function (NEGF) method. The physical
information is usually stored in two propagators, i.e. the retarded,
$G^r$, and the lesser, $G^<$, Green's functions which are defined in
the real time axis. These Green's functions can be related to the
contour-ordered Green's function  $G_C(t, t')=-i\hbar^{-1}\langle
T_C\{\psi(t)\psi^\dag(t')\}\rangle$, along the Keldysh contour, by
the Langreth rules.\cite{HaugJauho} The retarded Green's function
can be found from the Dyson equation in the time-domain as
\begin{eqnarray}\label{Grt}
  \left( i\hbar \frac{\partial}{\partial
  t}-\epsilon_d(t)\right)G^r(t, t')=\hbar\delta(t-t')+\int
  dt''\Sigma^r(t,t'')G^r(t'', t').
\end{eqnarray}

The lesser Green's function is given by the Keldysh equation
\begin{eqnarray}\label{Glt}
G^<(t, t')=\int dt_1 dt_2G^r(t, t_1)\Sigma^<(t_1, t_2)G^a(t_2, t'),
\end{eqnarray}
where $G^a(t, t')=(G^r(t', t))^\dag$ is the advanced Green's
function and $\Sigma^{r/<}(t, t')$ is the double-time
retarded/lesser self-energy. Usually, we need another Green's
function, the greater Green's function $G^>(t, t')$ which can be
found by either using the Keldysh equation by replacing $\Sigma^{<}$
with the greater self-energy $\Sigma^>$ or from the identity:
\begin{displaymath}
G^>(t,t')-G^<(t,t')=G^r(t,t')-G^a(t,t').
\end{displaymath}

For general time-dependent Hamiltonian, the solution of the above
equations remains challenging. However, since our Hamiltonian is
periodic in time, the Floquet theorem can be used to simplify the
solution.\cite{PR138B979,PR304229} It is more desirable to work in
an enlarged Hilbert space which is defined by the direct product of
the local basis in space and the basis for the time-periodic
functions with the frequency $\Omega$. The basis set of time
periodic functions which can be chosen to have the form
$\exp(-ik\Omega t)$ with $k\in Z$ gives the so-called Floquet basis
to take into account the periodicity in a nonperturbative way.  We
will combine the Floquet theory and the Green's functions to
investigate the time-dependent transport properties of quantum dot
systems.

The following notations are defined and will be used in the rest of
the paper. The Fourier transform of a double time function $O(t,
t')$ is defined as
\begin{eqnarray}
O(t,\epsilon)=\int dt' e^{i\epsilon(t-t')}O(t,t').
\end{eqnarray}
The periodicity of the function $O$ indicates that it can be
decomposed into the Floquet basis as
\begin{eqnarray}
O(t,\epsilon)=\sum_k O_k(\epsilon) e^{-ik\Omega t},
\end{eqnarray}
where $O_k$ is the Fourier coefficient of the function
$O(t,\epsilon)$. Several authors have studied the time-dependent
transport by the Fourier coefficients of the Floquet-Green's
functions \cite{PRL90210602,PRB72245339,PRB72125349,PRB77165326},
where only single label of the Floquet states is needed. However, as
we will see later, it is more convenient to write this time-period
function $O(t,t')$ in a matrix form $\mathcal{O}$ in Floquet space.
In the following, all operators expressed in matrix form in Floquet
space will be denoted by calligraphic symbols.  The matrix element
$\mathcal{O}_{k,k'}$ are related to the Fourier coefficient by
\begin{eqnarray}
\mathcal{O}_{k,k'}(\epsilon)=O_{k-k'}(\epsilon+k'\hbar\Omega).
\end{eqnarray}

In the matrix form of Floquet-Green's function, it is more
convenient to work with the Floquet Hamiltonian which is defined by
\begin{eqnarray}
\mathcal{H}=H(t)-i\hbar\frac{\partial }{\partial t}.
\end{eqnarray}
Using the previously introduced notations, the double-time retarded
Green's function Eq.(\ref{Grt}) can be rewritten in Floquet space in
a compact form as the resolvent of the Floquet-Hamiltonian
\cite{JPCM20085224} as
\begin{eqnarray}\label{FGR}
\mathcal{G}^r(\epsilon)=\frac{1}{\epsilon-\mathcal{H}_{dot}(t)-\mathcal{E}^r(\epsilon)},
\end{eqnarray}
where $\mathcal{E}^r(\epsilon)$ is the double-time retarded
self-energy in Floquet space and $\mathcal{H}_{dot}$ is the
Floquet-Hamiltonian of the isolated dot. The derivation of the above
equation is given in the Appendix. Similarly, the advanced
Floquet-Green's function is given by
\begin{eqnarray}
\mathcal{G}^a(\epsilon)=[\mathcal{G}^r(\epsilon)]^\dag.
\end{eqnarray}
After Fourier transform, the lesser Floquet-Green's function which
in time-domain is given by Eq. (\ref{Glt}) can be rewritten in a
compact form as
\begin{eqnarray}\label{FGL}
\mathcal{G}^<(\epsilon)=\mathcal{G}^r(\epsilon)\mathcal{E}^<(\epsilon)\mathcal{G}^a(\epsilon),
\end{eqnarray}
with $\mathcal{E}^<$ represents the matrix form of the lesser
self-energy in Floquet space.

We can see that in the matrix form in Floquet space, both the
retarded and lesser Green's functions (Eq. (\ref{FGR}) and Eq.
(\ref{FGL})) have similar structures with the Keldysh equations for
the Green's function in \emph{stationary}
situations\cite{PRB505528}, though these Green's function in
time-domain are double-time functions. The main difference is that
the Floquet-Green's function is expressed in an enlarged Hilbert
space due to the periodicity condition.

\subsection{Noninteracting quantum dot}

For noninteracting dot, i.e. $U=0$, the Hamiltonian of the system
can be given with quadratic terms only. We assume the energy level
of the dot is modulated by a harmonic field. The dot Hamiltonian
takes the time-dependence $H_{dot}(t)=\sum_\sigma
(\epsilon_d+eV_g\cos(\Omega t)) d^\dag_\sigma d_\sigma$. The
corresponded Floquet Hamiltonian is then given by
\begin{eqnarray}
[\mathcal{H}_{dot}]_{k,k'}=\{\epsilon_d+k\hbar\Omega\}\delta_{k,k'}
+\frac{1}{2}V_g \delta_{k,k'\pm 1}.
\end{eqnarray}

In the presence of ac gate, the self-energies are due to the
coupling between the dot and the leads. For the dot-lead coupling,
the retarded and lesser self-energy which is given in time-domain as
\begin{eqnarray}\label{SEt}
\Sigma_{\alpha,\sigma}^r(t, t')&=&\sum_k V_{k\alpha}^2g^r_{k\alpha\sigma}(t, t'),\\
\Sigma_{\alpha,\sigma}^<(t, t')&=&\sum_k
V_{k\alpha}^2g^<_{k\alpha\sigma}(t, t'),
\end{eqnarray}
where $g$ represents the Green's function of the lead electrons at
equilibrium and $\alpha$ is the lead label.  The retarded and lesser
Green's functions of lead electrons are given, respectively, by
\begin{eqnarray}
g^r_{k\alpha\sigma}(t,
t')=-i\theta(t-t')\exp[-i\int^t_{t'}dt_1\epsilon_{k\alpha}(t_1)],
\end{eqnarray}
\begin{eqnarray}
g^<_{k\alpha\sigma}(t,t')=if_\alpha(\epsilon^0_{k\alpha})\exp[-i\int^t_{t'}
dt_1\epsilon_{k\alpha}(t_1)],
\end{eqnarray}
where $f_\alpha(\epsilon)$ is the Fermi-distribution function in the
$\alpha$  lead.

After Fourier transform and  some algebra, the self-energies (Eq.
(\ref{SEt})) can be rewritten in the Floquet basis as
\begin{eqnarray}
\mathcal{E}^{r/<}_{\alpha,\sigma;
k,k'}(\epsilon)=\Sigma^{r/<}_{\alpha\sigma}(\epsilon+k'\hbar\Omega)\delta_{k,
k'}.
\end{eqnarray}
Once we neglect the energy shift by the dot-lead coupling, the
static self-energy $\Sigma_\alpha$ can be given in the so-called
wide-band approximation by
\begin{eqnarray}
\Sigma^r_\alpha(\omega)&=&-\frac{i}{2}\Gamma_\alpha(\omega)\\
\Sigma^<_\alpha(\omega)&=&i\Gamma_\alpha(\omega)f_\alpha(\omega),
\end{eqnarray}
where $\Gamma_\alpha(\omega)=2\pi\sum_{k\in
\alpha}V^2_{k}\delta(\omega-\epsilon_{k})=2\pi\rho_\alpha(\omega)V^2$.

The transport properties can be given by the Green's functions and
their self-energies of the quantum dot. The expression for the
time-dependent current operator from left lead is related to the
time derivation of the total number operator in the left lead:
\begin{eqnarray}
J_L(t)=-e\frac{\partial N_L}{\partial t}=-\frac{ie}{\hbar}[H(t),
N_L],
\end{eqnarray}
with $N_L=\sum_{k\sigma}c^\dag_{k,L,\sigma}c_{k,L,\sigma}$. In terms
of the nonequilibrium Green's functions, the expectation value of
current can be found by
\begin{eqnarray}
\langle J_L(t)
\rangle=\frac{2e}{\hbar}\sum_\sigma\lim_{t'\rightarrow t} Re\{\int
dt_1[G^r_\sigma(t, t_1)\Sigma^<_{L,\sigma}(t_1, t')+G^<(t,
t_1)\Sigma_L^a(t_1, t')]\},
\end{eqnarray}
where $\Sigma^a_L$ is the advanced self-energy due to the coupling
to the left lead.

Due to the discrete nature of electrons, the current fluctuation is
nonzero even at zero temperature. This fluctuation, known as current
noise, contains information of electron correlations which is beyond
the reach of traditional conductance measurement. This current noise
is defined as\cite{PR3361}
\begin{eqnarray}
S_{LL}(t, t')=\frac{1}{2}\langle\{\Delta J_L(t), \Delta
J_L(t')\}\rangle,
\end{eqnarray}
where $\Delta J_L(t)=J_L(t)-\langle J_L(t)\rangle$ represents the
fluctuation of the current operator from the left lead from its
expectation value.

Using the the Floquet-Green's functions,  the time-dependent current
can be rewritten in a compact form with similar structure of the
time-independent Meir-Wingreen current formula as
\begin{eqnarray}
\langle J_L(t)\rangle =\frac{4e}{\hbar}Re\{\int
\frac{d\omega}{2\pi}[\mathcal{G}^r\mathcal{E}^<_L
+\mathcal{G}^<\mathcal{E}^a_L]_{k,0}(\omega) e^{-ik\Omega t}\},
\end{eqnarray}
where $\mathcal{E}^a_L$ and $\mathcal{E}^<_L$ are the advanced and
lesser self-energy in Floquet space due to the coupling to the left
leads. To find the time-averaged current over one period, one can
simply set $k=0$ in the above formula.

Inserting the current operator in the definition of the noise
definition and after some algebra, the time-averaged current
fluctuation at zero frequency can be given in the matrix form of the
Floquet-Green's functions and self-energies as
\begin{eqnarray}\label{SN}
\hat S_{LL}(\omega=0)&=&\frac{2e^2}{h}[\int d\epsilon \mathcal{E}_L^>\mathcal{G}^<+\mathcal{G}^>\mathcal{E}^<_L\\
\nonumber
&&-(\mathcal{G}^r\mathcal{E}^>_L+\mathcal{G}^>\mathcal{E}_L^a)(\mathcal{G}^r\mathcal{E}^<_L
+\mathcal{G}^<\mathcal{E}^a_L)\\ \nonumber
&&+\mathcal{G}^>(\mathcal{E}^r_L\mathcal{G}^r\mathcal{E}^<_L+\mathcal{E}^r_L\mathcal{G}^<\mathcal{E}^a_L
+\mathcal{E}^<_L\mathcal{G}^a\mathcal{E}_L^a)\\ \nonumber
&&+(\mathcal{E}^r_L\mathcal{G}^r\mathcal{E}^>_L+\mathcal{E}^r_L\mathcal{G}^>\mathcal{E}^a_L
+\mathcal{E}^>_L\mathcal{G}^a\mathcal{E}^a_L)\mathcal{G}^<\\
\nonumber
&&-(\mathcal{E}^r_L\mathcal{G}^>+\mathcal{E}^>_L\mathcal{G}^a)(\mathcal{E}_L^r\mathcal{G}^<
+\mathcal{E}^<_L\mathcal{G}^a)+h.c.]_{0,0}.
\end{eqnarray}

The above noise formula is exact for noninteracting dot  where the
Hamiltonian of the system is quadratic so that the Wick theorem can
be applied. In the presence of finite Coulomb interaction, we have
quartic terms. A direct application of Wick theorem is no longer
possible. A full diagrammatic expansion is needed to reach the
desired formula. However, such task remains formidable since
multi-particle Green's functions are required. In order to make use
of the power of Wick theorem, one has to rely on approximations such
as the  mean field approximation or slave boson techniques to
rewrite the Hamiltonian in quadratic form.

\subsection{Infinite-$U$ Kondo dot: Slave-Boson NCA}

For infinite-$U$, a well established tool to study the transport
properties in the Kondo regime is the
NCA\cite{PRB362036,PRB4913929,PRB4911040,PRB67245307}. The NCA is
the lowest order conserving approximation. It is widely used to
investigate the properties of Kondo dot at equilibrium or out of
equilibrium. Using NCA, the self-energies of the pseudo-particles
are given in the time-domain as
\begin{eqnarray}
\Xi^r_\sigma(t, t')&=&iV^2\sum_{k\alpha} g^>_{k\alpha\sigma}(t, t')B^r(t,t');\\
\Pi^r(t,t')&=&-iV^2\sum_{k\alpha\sigma}D^r(t, t')g^<_{k\alpha\sigma}(t',t);\\
\Xi^<_{\sigma}(t, t')&=&iV^2 \sum_{k\alpha\sigma} g^<_{k\alpha\sigma}(t, t')B^<(t, t');\\
\Pi^<(t,t')&=&-iV^2\sum_{k\alpha\sigma}D^<(t,
t')g^>_{k\alpha\sigma}(t',t),
\end{eqnarray}
where $g^>$ is the greater Green's function of electrons in leads.
\begin{eqnarray}
g^>_{k\alpha\sigma}(t,t')=-i(1-f_\alpha(\epsilon_{k\alpha\sigma}^0)\exp[-i\int^t_{t'}dt_1\epsilon_{k\alpha\sigma}(t_1)],
\end{eqnarray}
$\Xi$ is the self-energy of the pseudo-fermion while $\Pi$
represents the self-energy for the pseudo-boson Green's functions.

The double time Green's functions for the slave particles are
defined by:
\begin{eqnarray}
iD_\sigma(t,t')=\langle T_c f_\sigma(t) f^\dag_\sigma(t')\rangle;\\
iB(t, t')=\langle T_c b(t) b^\dag(t')\rangle.
\end{eqnarray}

The retarded and lesser Green's functions of the pseudo-boson and
pseudo-fermion, together with their self-energies  are obtained from
a set of self-consistent equations. The Dyson equations for the
retarded and lesser Green's functions in time-domain are given
by\cite{PRB67245307}:
\begin{eqnarray}
(i\frac{\partial}{\partial t}-\epsilon_\sigma)D^r_\sigma(t,
t')=\delta(t, t')+\int dt_1\Xi^r_\sigma(t, t_1)D^r_\sigma(t_1, t');
\end{eqnarray}
\begin{eqnarray}
i\frac{\partial}{\partial t}B^r(t, t')=\delta(t, t')+\int dt_1
\Pi^r(t, t_1) B^r(t_1, t');
\end{eqnarray}
\begin{eqnarray}
(i\frac{\partial}{\partial t}-\epsilon_d)D^<_\sigma(t, t')=\int
dt_1[ \Xi_\sigma^r(t, t_1)D^<_\sigma(t_1,t')
+\Xi^<_\sigma(t,t_1)D^a(t_1, t')];
\end{eqnarray}
\begin{eqnarray}
i\frac{\partial}{\partial t}B^<(t, t')=\int dt_1[\Pi^r(t,
t_1)B^<(t_1, t')+\Pi^<(t, t_1)B^a(t_1,t')].
\end{eqnarray}
The greater and advanced Green's functions are given in NCA by:
\begin{eqnarray}
D^>_\sigma(t,t')&&=i[D^r(t, t')-D^a(t, t')];\\
B^>(t,t')&&=i[B^r(t,t')-B^a(t,t')];\\
D^a(t, t')&&=[D^r(t',t)]^*;\\
B^a(t,t')&&=[B^r(t',t)]^*.
\end{eqnarray}

Making use of the periodicity of our problem, the above equations
can again be rewritten in matrix form in an enlarged space. The
Floquet-Green's function of the pseudo-fermion and pseudo-boson can
be given in a compact form in matrix form as:
\begin{eqnarray}
\mathcal{D}^r_\sigma(\epsilon)&&=\frac{1}{\epsilon-\mathcal{H}_{f}-\mathcal{K}^r_\sigma(\epsilon)};\\
\mathcal{B}^r(\epsilon)&&=\frac{1}{\epsilon-\mathcal{H}_b-\mathcal{P}^r(\epsilon)};\\
\mathcal{D}^<_\sigma&&=\mathcal{D}^r_\sigma\mathcal{K}^<_\sigma\mathcal{D}^a_\sigma;\\
\mathcal{B}^<&&=\mathcal{B}^r\mathcal{P}^<\mathcal{B}^a,
\end{eqnarray}
where $\mathcal{H}_f$ and $\mathcal{H}_b$ are the Floquet
Hamiltonian of the pseudo-fermion and pseudo-boson, respectively.
$\mathcal{D}$ ($\mathcal{B}$) is the Floquet-Green's function of the
pseudo-fermion (pseudo-boson) in the matrix form. Their
corresponding self-energies in the matrix form are denoted as
$\mathcal{K}$ and $\mathcal{P}$, respectively.  The matrix element
of the self-energies for pseudo-particles within NCA is given in
Floquet space by:
\begin{eqnarray}
\mathcal{K}^r_{\sigma;k_1,k_2}(\omega)&=&\sum_{k'}\int\frac{d\omega'}{2\pi}
\mathcal{E}^>_{\sigma;k_1,k'}(\omega'-k'\hbar\Omega)\mathcal{B}^r_{k',k_2}(\omega-\omega');\\
\mathcal{K}^<_{\sigma;k_1,k_2}(\omega)&=&\sum_{k'}\int\frac{d\omega'}{2\pi}\mathcal{E}^<_{\sigma;k_1,k'}
(\omega'-k'\hbar\Omega)\mathcal{B}^<_{k',k_2}(\omega-\omega');\\
\mathcal{P}^r_{k_1,k_2}(\omega)&=&-\sum_{\sigma,k'}\int\frac{d\omega'}{2\pi}
\mathcal{D}^r_{\sigma;k_1,k'}(\omega+\omega')\mathcal{E}^<_{\sigma;k',k_2}(\omega'-k_2\hbar\Omega);\\
\mathcal{P}^<_{k_1,k_2}(\omega)&=&-\sum_{\sigma,k'}\int\frac{d\omega'}{2\pi}
\mathcal{D}^<_{\sigma;k_1,k'}(\omega+\omega')\mathcal{E}^>_{\sigma;k',k_2}(\omega'-k_2\hbar\Omega).\\
\end{eqnarray}
These equations can be solved in a self-consistent way. At each
iteration in the numerical calculations, the constraint condition
Eq. \ref{Cost} is numerically checked as
\begin{eqnarray}
\int \frac{i}{2\pi}d\epsilon [\mathcal{B}^<(\epsilon)-\sum_\sigma
\mathcal{D}_\sigma^<(\epsilon)]=1.
\end{eqnarray}

From the time-derivation of the total electron number in the left
lead, the time-averaged current formula can be found with the help
of Floquet-Green's functions as
\begin{eqnarray}
J=\frac{2e}{\hbar}\sum_\sigma\int\frac{d\epsilon}{2\pi}\{[\mathcal{D}^r_\sigma\mathcal{K}^<_{\sigma}
+\mathcal{D}^<_\sigma\mathcal{K}^a_\sigma](\omega)\}_{0,0}.
\end{eqnarray}

Due to the strong correlations, an exact formula for noise in Kondo
regime is formidable, if it exists. One has to make some
approximation. Along the line for the derivation of Eq. (\ref{SN}),
a noise expression in the slave particle Floquet-Green's functions
within NCA can be obtained as
\begin{eqnarray}
  \hat S_{LL}(\omega=0)&=&\frac{2e^2}{h}[\int d\epsilon \mathcal{K}_L^>\mathcal{D}^<+\mathcal{D}^>\mathcal{K}^<_L\\
\nonumber
&&-(\mathcal{D}^r\mathcal{K}^>_L+\mathcal{D}^>\mathcal{K}_L^a)(\mathcal{D}^r\mathcal{K}^<_L
+\mathcal{D}^<\mathcal{K}^a_L)\\ \nonumber
&&+\mathcal{D}^>(\mathcal{K}^r_L\mathcal{D}^r\mathcal{K}^<_L+\mathcal{K}^r_L\mathcal{D}^<\mathcal{K}^a_L
+\mathcal{K}^<_L\mathcal{D}^a\mathcal{K}_L^a)\\ \nonumber
&&+(\mathcal{K}^r_L\mathcal{D}^r\mathcal{K}^>_L+\mathcal{K}^r_L\mathcal{D}^>\mathcal{K}^a_L
+\mathcal{K}^>_L\mathcal{D}^a\mathcal{K}^a_L)\mathcal{D}^<\\
\nonumber
&&-(\mathcal{K}^r_L\mathcal{D}^>+\mathcal{K}^>_L\mathcal{D}^a)(\mathcal{K}_L^r\mathcal{D}^<
+\mathcal{K}^<_L\mathcal{D}^a)+h.c.]_{0,0},
\end{eqnarray}
In the above derivations, we have, following Meir \emph{et
al.}\cite{PRL88116802}, neglected the vertex correction when
decoupling the correlation functions. One merit of this
approximation is that it can recover the zero-voltage
fluctuation-dissipation theorem\cite{PRL88116802}. This merit has
been tested in our numerical results.  In the following, we will use
the above noise formula to investigate the noise properties of ac
driving Kondo dot.

\section{Numerical Results and Discussion}

In this section, we apply the Floquet-Green's function approach with
infinite-$U$ NCA to calculate the LDOS and noise properties of
ac-driving Kondo dot. In the following calculations, we take the
Lorentzian shape LDOS of the lead. The coupling between the dot and
lead is then given by
\begin{eqnarray}
\Gamma_\alpha(\omega)=\Gamma_0\frac{D^2}{(\omega-\mu_\alpha)^2+D^2},
\end{eqnarray}
where $D$ is the half bandwidth of the lead, $\mu_\alpha$ is the
chemical potential. We will take $\Gamma_0=1$ as the energy unit in
the following discussions. For the sake of simplicity, we use
$\hbar=e=k_B=1$. Throughout calculation, we have fixed
$\epsilon_d^0=-2$ without other statement. The half bandwidth of the
lead is $D=25$. The Kondo temperature for infinite $U$ can be
estimated from $T_K=\frac{D}{\sqrt{2\pi
|\epsilon_d^0|}}\exp(-\pi|\epsilon_d^0|)\approx 0.013$. The
temperature is chosen as $T=0.01$ without other statement which is
below the estimated Kondo temperature and not in the pathology
regime of NCA.

\subsection{Time-averaged LDOS of Kondo dot}

One fingerprint of the Kondo effect is the sharp Kondo peak of the
LDOS at the Fermi energy. In the following, we study the
time-averaged LDOS of quantum dot which is driven by  an ac gate
voltage.  The LDOS of the Kondo dot is related to the retarded
Green's function of the dot electrons which can be found from
\begin{eqnarray}
G^r_\sigma(t,t')=i(D^r_\sigma(t,t')B^<(t',t)+D^<_\sigma(t,t')B^a(t',t)).
\end{eqnarray}
At equilibrium, the retarded Green's function after Fourier
transform is independent of time. When the system is driving by ac
field, the retarded Green's function after Fourier transform is
time-dependent and can be given in Floquet space as
\begin{eqnarray}
[\mathcal{G}^r_\sigma(\omega)]_{k_1,k_2}=i\int\frac{d\omega'}{2\pi}[\mathcal{D}^r_{\sigma;k_1,
k'}(\omega+\omega')\mathcal{B}^<_{k',k_2}(\omega'-k_2\hbar\Omega)+\mathcal{D}_{\sigma;k_1,k'}^<(\omega+\omega')
\mathcal{B}^a_{k',k_2}(\omega'-k_2\hbar\Omega)].
\end{eqnarray}

The time-averaged LDOS can then be easily obtained by
\begin{eqnarray}
  {\rho}(\omega)=-\frac{1}{\pi}
  \mathrm{Im}[\mathcal{G}^r(\omega)]_{0,0}.
\end{eqnarray}

In Fig. 1, we show the calculated LDOS ${\rho}(\omega)$ as a
function of energy $\omega$ of the quantum dot modulated by ac gate
via $\epsilon_d(t)=\epsilon_d^0+eV_{ac}\cos(\Omega t)$. Different ac
frequencies are used in our calculated as indicated in the figure.
The ratio between the ac strength and ac frequency is fixed at
$\Omega/V_{ac}=1$. From Fig. 1, we can see that the ac gate voltage
can modify the time-averaged LDOS. However, the Kondo peak is not
suppressed monotonically by increasing ac frequency (ac strength).
The evolution of the Kondo peak with increasing ac field is enlarged
in the inset of Fig. 1. For low ac field (ac frequency and ac
strength), for example, $\Omega = 1\ T_K$, the time-averaged LDOS is
almost identical with the equilibrium LDOS. The LDOS against ac
field remains robust in the numerical results as up to an external
ac frequency of $20\ T_K$. This value is much larger than the width
of the Kondo peak which can be estimated by its Kondo temperature
$T_K$.  From Fig. 1, by increasing the ac frequency from $1\ T_K$ to
$20 \ T_K$, one can find that the broad peak around $\epsilon_d^0$
is only slightly lowered while the Kondo peak is almost unchanged.
For these low frequencies, the time-averaged LDOS largely resembles
the averaged LDOS for quantum dot at equilibrium with time-dependent
energy level over a period of the ac modulation. The observed robust
of LDOS against not-strong ac field agrees with the findings in Ref.
\cite{PRB612146} where the Kondo dot is found to be not sensitive to
the weak ac field. However, when the ac field is strong, the
adiabatic picture is no longer valid. Significant suppression of the
Kondo peak is observed in Fig. 1 for $\Omega=V_{ac}=50\ T_K$ and
$100\ T_K$. At the same time, the width of the Kondo peak is
broadened.  For very large ac field (for $\Omega=50\ T_K$ and $100\
T_K$ in Fig. 1), the numerical results clearly show the inelastic
photon-assisted processes. One can observe in Fig. 1 that replica of
Kondo peak appears in the LDOS with a distance to the Kondo peak
equals the ac frequency.

\subsection{Photon-assisted noise for quantum dot}
For a non-interacting quantum dot, analysis based on the scattering
approach has shown that the photo-assisted noise as a function of
the dc bias displays cusps at integer $\Omega/V$\cite{PRL72538}. The
derivation of the noise then gives staircase behavior. Such
singularities can be attributed to the change of the distribution of
the transmitted charge by ac field. Here, we calculate the
photon-assisted noise through a noninteracting quantum dot with ac
gate by the Floquet-Green's function method and noise formula Eq.
(\ref{SN}). Identical numerical results are also obtained by the
noise formula presented in Ref. \cite{PRL90210602} which is valid
for ac transport in the absence of Coulomb interaction. In Fig. 2,
we present the time-averaged noise power (in unit
$\frac{e^2\Gamma_0}{h}$) as a function of dc bias voltage $V$ of ac
gate modulated noninteracting dots for different dot energy levels.
In the calculation, the temperature is fixed at $T=0.002$. The ac
frequency is chosen as $\Omega=0.2$ and the ac strength is
$V_{ac}=\Omega$. Since the magnitude of noise varies a lot with the
energy level position, we have rescaled the $y$ axis to make the
time-averaged noise $S(V)$ behavior at $V/\Omega=1$ more clearly. We
can see from Fig. 2 that the appearance of cusp in the
photon-assisted noise depends on the position of the energy level of
the quantum dot. When $\epsilon_d=0$, i.e. the dot is in resonance,
the noise shows an obvious cusp at $V=\Omega$. The cusp is ascribed
to the non-adiabatic photon-assisted tunneling\cite{PRL72538}. Since
we have chosen a nonzero temperature, the sharpness of this cusp has
been smoothed. When we lower the energy level and tune the dot out
of resonance, the noise curves become smooth around
$eV_{dc}=\hbar\Omega_{ac}$. These results indicate that the
singularity of $dS/dV$ for noninteracting dot in the presence of ac
gate modulation is most pronounce when the dot is situated at
resonance. However, when the dot is far from resonance, the noise
behaviors become smooth and it is hard for experiments to detect the
noise singularity behavior.

We have shown that the non-interacting \emph{resonant} dot can
display noise singularity when the dot is modulated by ac gate. An
intuitive analogous between the noninteracting resonant dot and the
Kondo resonance, where both resonances can reach the conductance
quanta ($2e^2/h$ with spin degeneracy), may lead to the naive
conclusion that the noise of Kondo dot will show singularity in the
presence of ac gate since there is Kondo resonance at Fermi energy.
However, this statement is not correct, at least in the strongly
correlated infinite-$U$ regime, as shown in our numerical results
Fig. 3.  In Fig. 3, we show the noise properties of the Kondo dot
with different ac frequency as a function of the applied dc voltage.
Since we are interested in the noise behavior at $V/\Omega=1$, the
$x$ axis is rescaled by the ac frequency to have a better view at
$V/\Omega=1$ (note different values of $\Omega$ in different
curves). In Fig. 3, no singularity is observed at $V/\Omega=1$ for
all the ac parameters. For lowest ac frequency ($\Omega=1\ T_K$),
this is in agree with our previous results obtained by slave-boson
mean-field approximation\cite{PRB77233307}. The noise at weak ac
field is almost identical to the numerical results without ac field.
It is interesting to see that both the slave-boson mean-field
approximation and noncrossing approximation, which are valid in
their respective parameter space, show that the noise of Kondo dot
remains almost unaffected by the ac gate. The Kondo dot behaves as
if the ac gate is effectively screened to induce singularity in the
noise behavior.  Comparing with the noninteracting resonance
tunneling, the disappear of singularity behavior in noise of Kondo
dot may be ascribed to the fact that the electrons through the Kondo
resonance are not \emph{directly} modulated by the ac field.
Although the energy level of the quantum dot $\epsilon_d^0$ is
periodically modulated by the ac gate, the Kondo peak is not
directly driven by the ac field. For weak ac field where the ac
frequency or ac strength are not in the order of the coupling
strength or the energy level, the ac gate voltage is too small to
drive the dot out of the Kondo regime. Strong correlation can still
give rise to the sharp Kondo peak at Fermi energy. From the
numerical results of LDOS, we can see that ac modulation of Kondo
peak is much smaller than the applied ac gate which modulated the
energy level of the dot. When the electrons are tunneling though the
dot via the Kondo resonance, their transport behavior is not
significantly modulated by the ac field. The influence of the
time-periodicity of the Kondo peak to the electrons transport
through the dot is negligibly small to show any singularities in the
noise behavior.  However, when the ac strength is strong enough to
be comparable to $\epsilon_d^0$ or $\Gamma$, for example $\Omega=20\
T_K$ in Fig. 1 and 3, the quantum dot can sometimes be driven out of
the Kondo regime by the ac field. The magnitude of the Kondo peak
can then be significantly suppressed due to the strong ac field. As
a consequence, the conductance declines drastically and deviates
from the unitary limit. In these situations, the Kondo peak is then
significantly suppressed and electron transmission probability
becomes too small to show significant noise singularity.

Since the cusp of noise behavior can be smoothed by
finite temperature, it is important to exclude the possibility that
the smooth noise behavior shown in Fig. 3 is due to the finite
temperature ($T=0.01<T_K$) in our numerical simulation. In Fig. 4,
we plot the noise behavior as a function of dc bias voltage for the
dot with different temperature. The ac frequency and ac strength is
fixed at $\Omega=V_{ac}=1\ T_K$. The temperatures are chosen to be
0.01, 0.005 and 0.001, respectively. From Fig. 4, we can see that
even for the lowest temperature $(T=0.001)$ which is much lower than
$T_K$, no cusp is observed at $V/\Omega=1$ in the noise behavior.
Together with the numerical results presented in Ref.
\onlinecite{PRB77233307} where no singularity is observed at zero
temperature, the disappearance of cups in noise behavior can not be
ascribed to the finite temperature. It is a generic feature of the
strongly-correlated nature of Kondo peak.

Our numerical results together with those of Ref. \cite{PRB77233307}
which are obtained with the help of the slave-boson technique have
clearly shown that the strongly-correlated Kondo resonance can
display different noise behavior as compared with the noninteracting
dot. Usually, both the noninteracting resonant tunneling and Kondo
resonance will lead to similar results in conductance measurement.
Therefore, it is possible to make use of the distinct
photon-assisted noise behavior to tell apart whether the resonant
transport is via noninteracting resonance or strongly-correlated
Kondo resonance. Such measurement has been in the reach of present
technology. The recent experiment work\cite{PRL802437,PRL100026601}
has shown the possibility to conduct the noise measurement of
nanostructures with ac field. The ability to distinguish the origin
of the resonant transport may be helpful to answer the controversy
about origin of the 0.7 anomaly in quantum point contacts, where the
zero bias anomaly is due to whether the formation of strongly
correlated Kondo peak\cite{PRL89196802} or single-particle resonant
peak\cite{PRB75045346,PRB79153303} is still under debating. We wish
the experiment of the photon-assisted noise measurement can be used
to reveal the underlying physics of 0.7 anomaly.

\section{Conclusions}
In conclusion, we have presented a Floquet-Green's function
formalism to investigate the noise properties of quantum dots in the
ac-Kondo regime. The Coulomb correlation is taken into account by
the infinite-$U$ NCA. In principle, the ac effect can be considered
non-perturbatively via the infinite Floquet states.  Our results
show that the Kondo peak is robust against weak ac gate modulation.
Significant suppression of the Kondo peak can be observed when the
ac gate field becomes strong. The photon-assisted noise of Kondo
dots as a function of dc voltage does not show singularities which
are expected for noninteracting resonant quantum dot at integer
$eV/\hbar\Omega$, where $V$ is the dc bias voltage and $\Omega$ is
the ac frequency. These results suggest that we can tell resonant
transport apart from the Kondo resonance by photon-assisted noise
measurement, which can not be distinguished via the conductance
measurement.

\begin{acknowledgements}
The authors are grateful to Prof. JongBae Hong for helpful
discussions on the Kondo physics. Correspondence from Prof. Yigal
Meir is appreciated.
\end{acknowledgements}

\newpage

\appendix{Appendix}

In this appendix, we outline the derivation to reach the resolvent
form of the Floquet-Green's function. It resembles much to the
resolvent form of the Green's function at steady state. The starting
point is the definition of Green's function in time domain
\begin{eqnarray}\label{1}
  (i\hbar\frac{\partial}{\partial t}-H(t))G(t,t')=\delta(t-t')
\end{eqnarray}

Make Fourier transform to the two side of Eq. [\ref{1}]
\begin{eqnarray}\label{2}
  \int e^{i\omega(t-t')}dt'(i\hbar\frac{\partial}{\partial
  t}-H(t))G(t,t')=1
\end{eqnarray}

The Fourier transform of the Green's function is defined as
\begin{eqnarray}\label{3}
  G(t,t')=\int \frac{d\omega'}{2\pi}e^{-\omega'(t-t')}G(t,\omega').
\end{eqnarray}

Inserting Eq. (\ref{3}) into (\ref{2}), we have
\begin{eqnarray}
  &&\int e^{i\omega(t-t')}dt'(i\hbar\frac{\partial}{\partial
  t}-H(t))\int \frac{d\omega'}{2\pi}
  e^{-\omega'(t-t')}G(t,\omega')\\ \nonumber
  =&&\int \int
  e^{i(\omega-\omega')(t-t')}(\hbar\omega'+i\hbar\frac{\partial}{\partial
  t}-H(t)) G(t, \omega') dt' \frac{d\omega'}{2\pi}\\ \nonumber
  =&&\int \delta(\omega-\omega')d\omega'
  (\hbar\omega'-\mathcal{H}_F(t)) G(t, \omega')\\ \nonumber
  =&& [\hbar\omega-\mathcal{H}_F(t)]G(t,\omega)=\bm{1}
\end{eqnarray}
where we have introduced the Floquet Hamiltonian
$\mathcal{H}_F(t)=H(t)-i\hbar\frac{\partial}{\partial t}$.

Therefore, we arrive at the resolvent form of the Green's function
\begin{eqnarray}
  G(t, \omega)=\frac{1}{\hbar\omega-\mathcal{H}_F(t)}
\end{eqnarray}

\clearpage
\begin{figure}
  \includegraphics{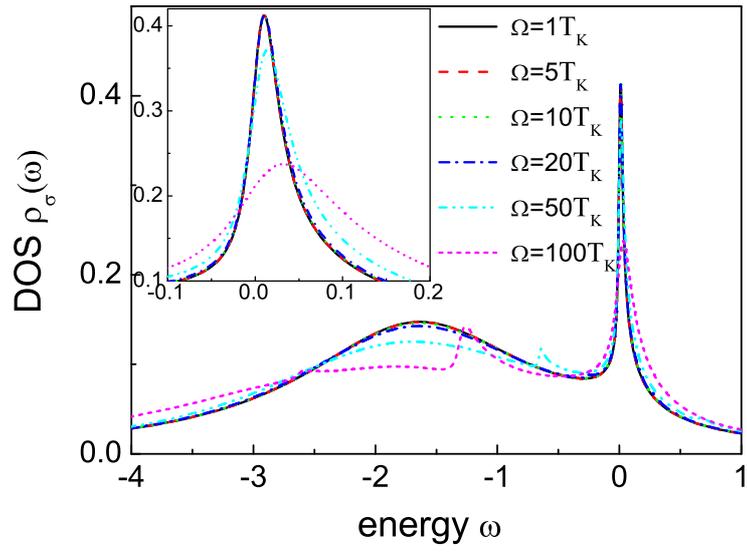}
  \caption{LDOS of Kondo quantum dots in the presence of ac gate modulation for different ac frequencies. The ac strength
  is fixed at $eV_{ac}/\hbar\Omega=1$.}
\end{figure}

\newpage
\clearpage

\begin{figure}
  \includegraphics{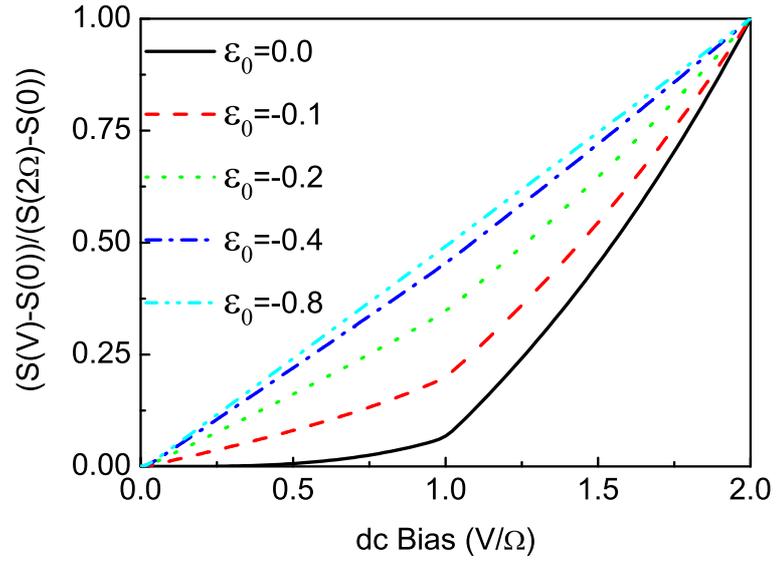}
  \caption{Noise of noninteracting quantum dot with ac gate field as a function of the applied
  dc bias for different values of energy level. }
\end{figure}

\newpage
\clearpage
\begin{figure}
  \includegraphics{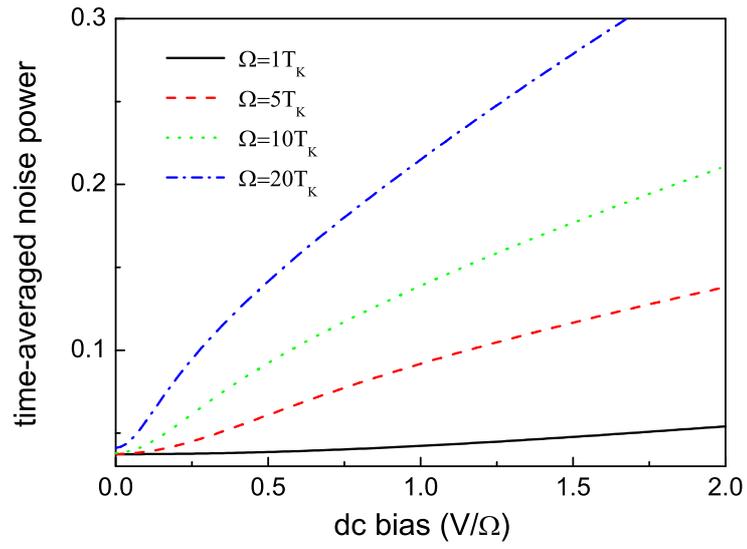}
  \caption{Noise of Kondo dot as a function of applied dc bias at fixed $eV_{ac}/\hbar\Omega=1$ and for
  four different values of frequency (in unit of the Kondo temperature). No singularities are observed at
  $eV_{dc}/\hbar\Omega=1$ for the different ac field parameters.}
\end{figure}

\newpage
\clearpage
\begin{figure}
  \includegraphics{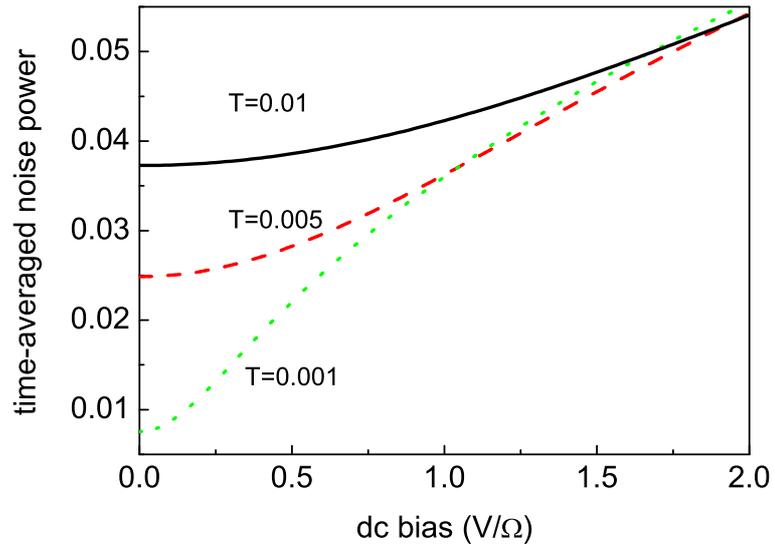}
  \caption{Noise of Kondo dot as a function of applied dc bias for
  different temperatures. The ac frequency and ac strength are fixed
  at $1\ T_K$.}
\end{figure}

\clearpage

\end{document}